\documentstyle[twocolumn,aps,eqsecnum,amssymb,epsf,epsfig,floats]{revtex}
\newcommand{\tr}{\text{tr}}
\newcommand{\probe}{\text{probe}}
\newcommand{\meas}{\text{meas}}
\newcommand{\qubit}{\text{qubit}}
\newcommand{\repeated}{\text{repeated}}
\newcommand{\overtwo}{\textstyle{1 \over 2}}
\newcommand{\overfour}{\textstyle{1 \over 4}}
\newcommand{\overeight}{\textstyle{1 \over 8}}
\newcommand{\pitwo}{\textstyle{\pi  \over 2}}
\newcommand{\roottwo}{\textstyle{1 \over {\sqrt 2}}}
\newcommand{\ket}[1]{{|\, #1\,\rangle}}
\newcommand{\kU}{{|\!\uparrow\,\rangle}}
\newcommand{\kD}{{|\!\downarrow\,\rangle}}
\newcommand{\kE}{{|\,\text{E}\,\rangle}}
\newcommand{\kW}{{|\,\text{W}\,\rangle}}
\newcommand{\kN}{{|\,\text{N}\,\rangle}}
\newcommand{\kS}{{|\,\text{S}\,\rangle}}
\newcommand{\bra}[1]{{\langle\, #1\, |}}
\newcommand{\bE}{{\langle\,\text{E}\, |}}
\newcommand{\bW}{{\langle\,\text{W}\, |}}
\newcommand{\bN}{{\langle\,\text{N}\, |}}
\newcommand{\bS}{{\langle\,\text{S}\, |}}
\newcommand{\PE}{P_{\text{E}}}
\newcommand{\PW}{P_{\text{W}}}
\newcommand{\PN}{P_{\text{N}}}
\newcommand{\PS}{P_{\text{S}}}
\newcommand{\szz}{\sigma _{zz}}
\newcommand{\szo}{\sigma _{z0}}
\begin{document}
\draft
\title{Measuring the entangled Bell and GHZ aspects using a single-qubit shuttle}
\author{Ulf Larsen\cite{byline}}
\address{Niels Bohr Institute, University of Copenhagen \\
Blegdamsvej 17, DK-2100 Copenhagen {\O}, Denmark }
\date{Dec. 30, 1998}
\maketitle
\begin{abstract}
A complete, non-demolition procedure is established for measuring 
multi-qubit entangled states, such as the Bell-states and the 
GHZ-states, which is essential in certain processes of quantum 
communication, computation, and teleportation.  No interaction between 
the individual parts of the entangled system, nor with any environment 
is required.  A small probe (e.g. a single qubit) takes care of all 
interaction with the system, and is used repeatedly.  The probe-qubit 
interaction is of the simplest form, and only this one type of 
interaction is required to perform a complete measurement.  The 
process may be divided into elementary local operations and 
interactions, taking place sequentially as the probe visits each of 
the qubits.  A shuttle mode is described, which may be repeated 
indefinitely.  By the quantum Zeno effect, the entangled states can be 
maintained until released in a predictable state.  This shuttle 
process is stable, and self-correcting, by virtue of the standard 
measurements performed repeatedly on the probe.
\end{abstract}
\section{Introduction}\label{section1}
Quantum entanglement is one of the most remarkable manifestations of 
the fundamental principles.  Consider two distinguishable systems, with 
quantum states $\ket{\psi _i}$ and $\ket{\phi _i}$, $i=1,2\cdots $. 
When these systems are independent, then their joint state is 
separable, as for instance $\ket{\psi _1} \otimes \ket{\phi _1}$, or 
$\ket{\psi _2} \otimes \ket{\phi _2}$.  However, according to the 
principle of superposition, the compound system may also exist in 
states that are not separable, such as
\[ \ket{\Psi } =c_1\ket{\psi _1} \otimes \ket{\phi _1}+c_2\ket{\psi _2} \otimes  \ket{\phi _2} +\cdots \]
where $c_1, c_2,\cdots $ are complex amplitudes. This possibility has given 
rise to many profound discussions, in particular concerning the EPR paradox and 
the Bell inequalities 
\cite{WheelerZurek83,Bell87}. It has been confirmed\textemdash beyond any reasonable 
doubt\textemdash that these entangled states are uniquely a quantum phenomenon 
\cite{Aspect82,Tittel98,Weihs98}.
\par 
There have been many proposals for preparing specific entangled states, 
starting with suitably initialized states, such as for instance
\cite{Phoenix93,Zukowski93,Cirac94,Gerry96,Zeilinger97,Bose98}, 
and this has been achieved in new ways, for Bell-states 
\cite{Braunstein92,Kwiat95,Hagley97,Turchette98} (notation, cf. Sect.~\ref{subsect:defs}) 
\begin{eqnarray*}
\ket{\Phi ^\pm } &=& \roottwo \bigl( \ket{\!\uparrow \uparrow} \pm \ket{\!\downarrow \downarrow} \bigr) \\
\ket{\Psi ^\pm } &=& \roottwo \bigl( \ket{\!\uparrow \downarrow} \pm \ket{\!\downarrow \uparrow} \bigr) 
\end{eqnarray*}
and for GHZ-states \cite{Greenberger89,Greenberger90,Mermin90,Bouwmeester98,Laflamme98}
\[ \ket{\Xi ^\pm } = \roottwo \bigl( \ket{\!\uparrow \uparrow \uparrow} \pm \ket{\!\downarrow \downarrow \downarrow} \bigr) \]
\par 
Recent experiments have demonstrated that it is feasible to design 
processes that depend in an essential way on handling such entangled 
states.  This includes quantum communication 
\cite{Mattle96}, quantum computation 
\cite{Monroe95,Turchette95,Jones98a,Jones98b,Chuang98a,Chuang98b,Cory98a}, and quantum teleportation 
\cite{Bouwmeester97,Boschi98,Pan98a,Furusawa98,Nielsen98}.  For 
instance, in densely coded quantum communication the receiving party 
must measure the Bell-states in order to retrieve the encoded information 
\cite{Bennett92}.  In quantum teleportation 
\cite{Bennett93} the executive step is also a measurement of the entangled Bell-states, or an EPR-state 
\cite{Vaidman94,Braunstein98}.  
This projects the entire system into the teleporting configuration, 
and, at the same time, acquires the data that must be transmitted by 
classical means.
\par 
In order to perform such a measurement one has to design a global 
experimental situation for the compound system.  For pairs of photons, 
interferometric methods can provide characteristic detection patterns, 
which more or less completely distinguish the entangled states from each other 
\cite{Bouwmeester97,Pan98a,Weinfurter94,Zeilinger94,Braunstein95,Michler96,Pan98b}. 
Unless interaction is possible a complete Bell-state measurement is not feasible 
\cite{Vaidman98,Lutkenhaus98}. In 
\cite{Boschi98}, where the entangled components are the 
polarization and momentum of a single photon, all the Bell-states can 
be distinguished by using polarizing beam-splitters 
\cite{Popescu95}.  
In this situation the entangled subsystems (i.e. 
polarization and momentum of a single photon) are made to interact 
directly with each other.  That is also the case in a recent NMR 
based teleportation experiment 
\cite{Nielsen98}.  In these procedures, 
what amounts to a controlled-NOT operation disentangles the Bell-states 
\cite{Barenco95,Bruss97}, and one can then measure the now 
separate subsystems.  The entangled state is destroyed in these types 
of measurement.
\par 
However, states of known entanglement are becoming a resource.  It 
will be desirable to retain the entangled system itself after the 
measurement, for subsequent processing\textemdash not merely the data 
acquired.
\par 
Also, from a more fundamental point of view, in order to establish 
entanglement as a standard physical observable, one really needs a 
non-demolition procedure.  That is, a procedure which leaves any of 
the properly entangled states invariant, while delivering unambiguous 
information about it.  A method of this nature has been proposed for the 
Bell-states and other entangled states of a similar structure in 
\cite{Vaidman94,Aharonov86}.  In this design, the measuring device 
itself consists of components that must be prepared beforehand into 
entangled states.  After these components have been placed at the 
relevant locations, an instantaneous non-local measurement can be 
performed, at least in principle.
\par 
The measurement procedure established in the present paper is 
different from the existing ones in several respects.  It is both 
simple and economical of resources that appear to remain scarce, at 
least within a foreseeable future.  The {\lq apparatus\rq} can be as 
small as a single-qubit probe, which is used repeatedly according to a 
predesigned shuttle schedule.  In principle, this allows a complete 
measurement of the entangled basis-states of any number of target 
qubits.  The necessary interactions have been reduced to bare 
essentials, so that only the most elementary operations are involved 
at every stage of the procedure. In particular, no entanglement is 
required at the outset. For example, it is possible to imagine a 
single $^{13}\text{C}$ nuclear spin probe, paying simultaneous 
attention to a number of different H qubits.
\par 
Since the method relies on a traveling probe, it conforms in a 
straightforward way with special relativity. On the other hand, this 
means that it is not capable of instantaneous measurement at 
space-like separation. However, it seems that it could be efficient 
in combination with entanglement swapping, in creating distant 
multi-qubit entangled states \cite{Zukowski93,Bose98}.
\par 
Sect.~\ref{section2} describes the principles of the procedure for 
measuring the Bell-states of two qubits.  Examples include both a 
spin~$1/2$, and a single-mode cavity-field probe.  
Sect.~\ref{section3} entends these methods to systems of arbitrary 
numbers of qubits, and provides an extensive analysis of the 
GHZ-states of three qubits.  For both cases a shuttle design is 
provided, which permits the measurement to be continued indefinitely, 
with a sustained flow of measurement data.  In effect, this could 
provide stable storage for entangled states, which can then (in 
principle, of course) be extracted with certainty at a predetermined 
time, using the acquired data.
\section{Measuring the Bell-aspect}\label{section2}
\subsection{Definitions}\label{subsect:defs}
Consider two qubits, labeled 1 and 2, and represented as 
distinguishable spin~$1/2$ systems with Pauli spin operators $\vec 
\sigma =(\sigma _x,\sigma _y,\sigma _z)$.  Let $\sigma _z \kU = \kU$ and $\sigma _z \kD =-\kD$, 
and write their joint states in the form
$$\ket{\!\uparrow \downarrow} \equiv \kU ^{(1)}\otimes \kD^{(2)}$$ 
The Bell-states form an orthonormal basis of (maximally) entangled two-qubit states
\begin{mathletters}
\begin{eqnarray}
\ket{\Psi ^\pm } &=& \roottwo \bigl( \ket{\!\uparrow \downarrow } \pm \ket{\!\downarrow \uparrow} \bigr) \label{x} \\
\ket{\Phi ^\pm } &=& \roottwo \bigl( \ket{\!\uparrow \uparrow } \pm \ket{\!\downarrow \downarrow} \bigr) \label{xx}
\end{eqnarray}
\end{mathletters}
These states are simultaneous eigenstates of the commuting operators
\begin{eqnarray}
\sigma _{xx}&\equiv &\sigma _x^{(1)}\otimes  \sigma _x^{(2)} \nonumber \\
\sigma _{yy}&\equiv &\sigma _y^{(1)}\otimes  \sigma _y^{(2)} \label{z} \\
\szz &\equiv &\sigma _z^{(1)}\otimes  \sigma _z^{(2)} \nonumber
\end{eqnarray}
with eigenvalues $\pm 1$. They are therefore also eigenstates of the 
{\lq Bell-operators\rq} \cite{Braunstein92,Clauser69}
$$B_{\text{CHSH}}=\sqrt{2}\bigl( \sigma _{xx} \pm \sigma _{yy} \bigr)$$ 
The eigenvalues $\pm 2\sqrt{2}$ signal a maximal violation 
of the corresponding Bell-inequalities, in the well-known way. The 
combination of this set of compatible observables, and their basis 
states, will be referred to as the {\lq Bell-aspect\rq} 
\cite{Larsen90}.
\par 
The eigenvalue of $\sigma _{xx}$ determines the {\lq e/o question\rq}, 
i.e. whether the state is even or odd (e/o) under global spin flip.  
The eigenvalue of $\szz $ determines the {\lq p/a question\rq}, 
whether the spins are parallel or antiparallel (p/a).  The 
corresponding projectors are
\begin{eqnarray}
\bar P_a=\overtwo \bigl( 1+a\sigma _{xx} \bigr)&,&\quad a=\pm 1 \label{v} \\
P_b=\overtwo \bigl( 1+b\szz  \bigr)&,&\quad b=\pm 1 \label{w}
\end{eqnarray}
Consequently, the Bell-aspect is defined by the one-dimen\-sional projectors
\begin{equation}
P_{ab}=\overfour \bigl( 1+a\sigma _{xx} \bigr) \bigl( 1+b\szz  \bigr)
\label{m}
\end{equation}
A complete measurement can therefore be done in two stages. Each 
stage consists of a partial measurement separately deciding the p/a 
and e/o questions. The relevant procedures must commute, as the 
projectors (\ref{v}) and (\ref{w}) do.
\par 
Let a bilateral rotation of the two spins, by $-\pi /2$ about the 
global y-axis, be written as
$$U_{yy}\equiv U_y\otimes  U_y,\quad U_y=\roottwo \bigl( {1+i\sigma _y} \bigr)$$
Then
$$\sigma _x=U_y^\dagger \sigma _zU_y,\quad {\left(U_y\right)}^2=i\sigma _y$$
Using the algebra of the Kronecker product one gets
\begin{eqnarray*}
\sigma _{xx}&=&U_{yy}^\dagger \szz U_{yy}\\
\sigma _{yy}&=&-{\left( {U_{yy}} \right)}^2=-{\left( {U_{yy}^\dagger } \right)}^2
\end{eqnarray*}
Therefore
\begin{equation}
P_{ab}=-\overfour \sigma _{yy}U_{yy}\bigl( 1+a\szz  \bigr) U_{yy}\bigl( 1+b\szz  \bigr)
\label{y}
\end{equation}
This suggests a simplified two-stage procedure, consisting of 
identical operations $U_{yy} \left( 1+a\szz  \right) $.  Here 
one first decides the p/a question, then one rotates the spins 
bilaterally.  With such a chain of operations, the measurement is 
reduced to the most economical form, with respect to the resources 
that will be needed to carry it out in practice.  In particular, only 
one type of unilateral rotation, $U_y$, is required.  The essential 
task is to perform the partial p/a measurement, but also here only a 
single procedure is necessary.  What is more, each partial measurement 
is a binary test.  It therefore requires no more than a single spin~$1/2$, 
repeatedly probing the qubits.
\subsection{Spin~$1/2$ probe}
Let the probe eigenstates in the xy-plane be denoted $\kE$, 
$\kW$, $\kN$, and $\kS$, i.e.
\begin{eqnarray*}
\sigma _x \kE = \kE &,&\quad \sigma _x \kW =-\kW \\
\sigma _y \kN = \kN &,&\quad \sigma _y \kS =-\kS \\
\sigma _z \kE = \kW &,&\quad \sigma _z \kW = \kE 
\end{eqnarray*}
The corresponding projectors are written
\begin{eqnarray}
\PE= \kE \bE&,&\quad \PW= \kW \bW \nonumber \\
\PN= \kN \bN&,&\quad \PS= \kS \bS
\label{l}
\end{eqnarray}
In the following, $\vec \sigma =(\sigma _x,\sigma _y,\sigma _z)$ (with 
no superscript) stands for the Pauli operators of this probe. The 
explicit $\otimes $ separates the probe and the qubit operators.
\par 
The interaction between a single qubit and the probe is taken to be 
of the simplest possible form
$$U_1=\exp (-i\theta \sigma _z\otimes  \sigma _z^{(1)}/ 2)$$
The parameter $\theta $ can be adjusted by controlling the length of 
time during which the probe interacts with qubit 1.  This type of 
interaction can be realized in different systems, including 
nuclear spins, where it is standard, and by dispersive 
Rydberg-atom/cavity-field interaction (described in the following, 
cf. Sect.~\ref{subsect:cav}).  
It allows the entire measurement process to be reduced to elementary 
operations.  This strategy is of course familiar from NMR work, for 
instance, which establishes its feasibility.
\par 
A similar interaction with qubit 2 gives
$$U_2=\exp (-i\theta \sigma _z\otimes  \sigma _z^{(2)}/ 2)$$
It does not matter if these interactions with the qubits take place 
simultaneously, or one after the other. All operators commute, and
$$U_2U_1=\exp (-i\theta \sigma _z\otimes  S_z)$$
where
$$S_z={\textstyle{1 \over 2}}\bigl( {\sigma _z^{(1)}+\sigma _z^{(2)}} \bigr)$$
Extend the definitions (\ref{z}) to include the unit operators, such that 
for instance
$$\szo \equiv \sigma _z^{(1)}\otimes  1^{(2)}$$
Then one can use $\sigma _{0z}=\szo \szz $ to write
\begin{equation}
S_z={\textstyle{1 \over 2}}\left( {\szo +\sigma _{0z}} 
\right)=\szo P_+=\sigma _{0z}P_+ \label{q}
\end{equation}
The projector is given by (\ref{w}). For $j=1,2,\ldots $
\begin{eqnarray*}
\left( {\sigma _z\otimes  \szo P_+} \right)^{2j}&=&1 \otimes  P_+ \\
\left( {\sigma _z\otimes  \szo P_+} \right)^{2j+1}&=&\left( 
{\sigma _z\otimes  \szo } \right)\left( {1 \otimes  P_+} \right) 
\end{eqnarray*}
Therefore
\begin{eqnarray}
U_2U_1&=&1 \otimes  1+\left( {\cos \theta -1} \right)\left( {1 \otimes  P_+} \right) \nonumber \\
&&-i\sin \theta  \left( {\sigma _z\otimes  \szo } \right)\left( {1 \otimes  P_+} \right) \nonumber \\
&=&1 \otimes  P_- \label{p}\\
&&+\left( {\cos \theta -i\sin \theta \; \sigma _z\otimes  \szo } \right)\left( {1 \otimes  P_+} \right) \nonumber
\end{eqnarray}
For $\theta =\pi / 2$ then
\begin{equation}
U_2U_1=1 \otimes  P_--i\sigma _z\otimes  \szo P_+ \label{B}
\end{equation}
Here $\sigma _z$ rotates the probe spin, while $\szo $ swaps 
the qubit Bell-states, e.g.
$$\szo \ket{\Phi ^\pm } = \ket{\Phi ^\mp } $$
Because of the $P_+$ projector, both of these operations take place 
only when the qubit spins are in the parallel subspace spanned by 
$\Phi ^\pm $.
\subsection{The partial measurement (p/a)}
Let the initial states of the probe and qubits be given by statistical 
operators (density matrices) $v$ and $w$, respectively.  Suppose the 
probe is initialized along the positive x-direction, so that the joint 
state is
$$\rho =v \otimes  w,\quad v=\PE$$
After the probe-qubit interactions one has
\begin{eqnarray*}
\rho '&=&U_2U_1 \; \rho \; U_1^\dagger U_2^\dagger \\
&=&\kE\bE\otimes  P_-wP_- \\
&&+\kW\bW\otimes \szo P_+wP_+\szo  \\
&&+i\,\kE\bW\otimes  P_-wP_+\szo  \\
&&-i\,\kW\bE\otimes \szo P_+wP_-
\end{eqnarray*}
One can now read the probe by any ordinary, local measurement, 
coupling it to a suitable apparatus, possibly including a large 
environment. As far as the probe and qubit degrees of freedom are 
concerned, this is equivalent to the operation
\begin{eqnarray}
\rho '_{\meas}&=&\left( {\PE\otimes  1} \right) \rho  \left( {\PE\otimes 1} \right) \nonumber \\
&&+\left( {\PW\otimes  1} \right) \rho  \left( {\PW\otimes  1} \right) \label{r}
\end{eqnarray}
That is, none of this should affect the qubit pair. Measuring the 
probe spin in the x-direction then gives
\begin{equation}
\rho '_{\meas}=\PE\otimes  P_-wP_-+\PW\otimes  \szo P_+wP_+\szo 
\label{s}
\end{equation}
This represents the acquisition of binary data, {\lq E\rq} or 
{\lq W\rq} corresponding to the distinct alternatives of the 
p/a test.  The (reduced) state of the qubit-pair is not affected by 
these interactions between the probe and the external apparatus
\begin{eqnarray*}
w'&=&\tr _{\probe }(\rho ')=\tr _{\probe }(\rho '_{\meas}) \\
&=&P_-wP_-+\szo P_+wP_+\szo 
\end{eqnarray*}
The statistics of the data is that, one obtains the item E a 
fraction of the times given by 
\begin{eqnarray*}
\tr \bigl( {(\PE \otimes 1) \rho ' (\PE \otimes  1)} \bigr)&=&\tr _{12}(P_-wP_-) \\
&=&\tr _{12}(wP_-) \equiv  p_-
\end{eqnarray*}
This equals the estimated probability, $p_-$, for anti-parallel spins 
in the initial qubit state $w$. Likewise, one gets W with a frequency
$$\tr _{12}(\szo P_+wP_+\szo )=\tr _{12}(wP_+) \equiv  p_+$$
\subsection{Probe motion}
It is also of interest to monitor the probe spin in the course of 
time during the interaction with the qubits. The reduced probe state 
at a time corresponding to $\theta $ is given by (using (\ref{p}))
\begin{eqnarray*}
v(\theta )&=&\tr _{12}(U_2U_1 \, \rho \, U_1^\dagger U_2^\dagger ) \\
&=&\left( {p_-+p_+\cos ^2\theta } \right) v+p_+\sin ^2\theta \; \sigma _zv\sigma _z \\
&&+i\sin \theta \cos \theta \; \langle S_z \rangle \left( {v\sigma _z-\sigma _zv} \right) 
\end{eqnarray*}
Here, according to (\ref{q}),
\begin{eqnarray*}
\langle S_z \rangle &=&\tr _{12}(w\szo P_+)  \\
&=&\bra{\Phi ^-} \: w \: \ket{\Phi ^+} +\bra{\Phi ^+} \: w \: \ket{\Phi ^-} \\
&=&\bra{\uparrow \uparrow \!} \: w \: \ket{\! \uparrow \uparrow} 
-\bra{\downarrow \downarrow \!} \: w \: \ket{\! \downarrow \downarrow}  
\end{eqnarray*}
This quantity is zero in all the Bell-states, but not in the triplet 
states where $S_z=\pm 1$. The probe spin polarization is
$$\vec m(\theta )=\tr _{\probe}\bigl( v(\theta )\,\vec \sigma \bigr) $$
This gives
\begin{eqnarray*}
m_x(\theta )&=&\bigl( {p_-+p_+\cos 2\theta } \bigr) \; m_x(0)-\sin 2\theta 
\; \langle S_z \rangle \; m_y(0) \\
m_y(\theta )&=&\bigl( {p_-+p_+\cos 2\theta } \bigr)  \; m_y(0)+\sin 2\theta 
\; \langle S_z \rangle \; m_x(0) \\
m_z(\theta )&=&m_z(0) 
\end{eqnarray*}
For a large ensemble this motion may be detectable, and would then 
provide information about $w$ in terms of the mean values $p_\pm $ and 
$\langle S_z \rangle $. If no probe measurement is made, then (disregarding 
decoherence) everything eventually returns to the initial state $\rho $. 
On the other hand, in the measurement situation, where one starts at 
$\vec m=(1,0,0)$, one gets instead at $\theta =\pi / 2$
$$v'=p_-\PE+p_+\PW$$
and
\begin{eqnarray*}
m_x(\pi / 2)&=&p_--p_+ \\
m_y(\pi / 2)&=&m_z(\pi / 2)=0
\end{eqnarray*}
Of course, a measurement on the probe at this stage removes its 
coherence with the qubit pair, as it should, and the probe-qubit 
system therefore cannot return to the initial state. The data thus 
acquired determines the initial state of each individual probe-qubit 
system in the subsequent evolution. In the case of W this means 
that the probe will be in the state $\PW$, and, at the same time, 
the qubit pair state is projected into the parallel subspace. This 
occurs with a statistical frequency equal to the probability $p_+$.
\subsection{Repeated measurement}
One can cancel the $\Phi ^\pm $ permutation due to $\szo $ by 
performing a repeated measurement, using the output as new input. 
Then, using (\ref{s}) as initial state, either
$$\PE\otimes  P_-wP_-\mapsto \PE\otimes  P_-wP_-+0$$
or
\begin{eqnarray*}
\PW &\otimes& \szo P_+wP_+\szo  \\
&\mapsto& 0+\PE\otimes \szo P_+\szo P_+wP_+\szo P_+\szo  \\
&=&\PE\otimes  P_+wP_+ 
\end{eqnarray*}
Assuming one starts the probe initially in $\PE$, this produces data 
EE and WE, respectively, and the probe always ends up in 
$\PE$. The qubit pair experiences a clean projection
\begin{equation}
\rho '_{\repeated}=\PE\otimes  \left( {P_-wP_-+P_+wP_+} \right)
\label{I}
\end{equation}
Although somewhat extravagant, this duplication of the data is of 
course a useful check, and the resulting state is the ideal input for 
the second stage in a complete measurement. It is essential that one 
perform the probe measurement at the end of both of the $U_2U_1$ interaction 
sequences. Another way to compensate for the unitary transformation by 
$\szo $ will be described in Sect.~\ref{subsect:comp}.
\begin{table}
\caption{Bell-state propagation during shuttle operation (phases 
omitted). The probe starts in the state $\PE$. \label{table1}}
\begin{tabular}{ccccccccc}
$ab$&input&$U_2U_1$&$U_{yy}$&$M_x$&$U_1U_2$&$U_{yy}$&$M_x$&flips \\
\tableline
$++$&$\Phi ^+$&$\Phi ^-$&$\Psi ^+$&W&$\Psi ^+$&$\Phi ^-$&W&10 \\
$+-$&$\Psi ^+$&$\Psi ^+$&$\Phi ^-$&E&$\Phi ^+$&$\Phi ^+$&W&01 \\
$-+$&$\Phi ^-$&$\Phi ^+$&$\Phi ^+$&W&$\Phi ^-$&$\Psi ^+$&E&11 \\
$--$&$\Psi ^-$&$\Psi ^-$&$\Psi ^-$&E&$\Psi ^-$&$\Psi ^-$&E&00 \\
\end{tabular}
\end{table}
\subsection{Second partial measurement (e/o)} 
According to (\ref{y}) a complete measurement consists of a sequence (reading 
from right to left and omitting the final $\sigma _{yy}$)
\begin{equation}
\left( {M_xU_{yy}U_2U_1} \right)\left( {M_xU_{yy}U_2U_1} \right)
\label{t}
\end{equation}
Here, $M_x$ stands for any (unspecified) interaction of the probe with its 
supporting apparatus, so that the outcome is the required probe 
measurement in the x-direction. There are many different ways to do 
this, and it is only necessary that the result is the measurement 
operation of (\ref{r}), as far as the probe\rq s state is concerned. 
The simplest presumably is a process like the one analyzed in 
\cite{Larsen88}. Such a dephasing has recently been demonstrated in NMR
teleportation 
\cite{Nielsen98}, taking advantage of an order-of-magnitude difference 
in the $T_{1}$ and $T_{2}$ time-scales on the 
$^{13}\text{C}$ spins.
\par 
The first measurement decides the p/a test, the 
second one the e/o test. The bilateral rotation $U_{yy}$ of the qubits 
commutes with any $M_x$. As mentioned, since the probe is projected into either 
$\ket{\text{E}}$ or $\ket{\text{W}}$, it is automatically initialized for the 
following measurement. After the second measurement the state has become 
(using (\ref{s}))
\begin{eqnarray*}
\rho ''_{\meas}&=&\PE\otimes \left( {U_{yy}P_-U_{yy}P_-} \right) w \left( { \text{do.} } \right)^\dagger \nonumber \\
&&+\PW\otimes \left( {U_{yy}\szo P_+U_{yy}P_-} \right) w \left( { \text{do.} } \right)^\dagger \nonumber \\
&&+\PW\otimes \left( {U_{yy}P_-U_{yy}\szo P_+} \right) w \left( { \text{do.} } \right)^\dagger \nonumber \\
&&+\PE\otimes \left( {U_{yy}\szo P_+U_{yy}\szo P_+}\right) w \left( { \text{do.} } \right)^\dagger 
\end{eqnarray*}
Here $U_{yy}$ causes the following permutation
$$U_{yy}\ket{\Psi ^+} = \ket{\Phi ^-} ,\quad U_{yy}\ket{\Phi ^-} =-\ket{\Psi ^+} $$ 
and leaves the other two Bell-states invariant.  The handling of the 
Bell-states is summarized in Table~\ref{table1}.  It is understood 
that, permutations between the basis states within the Bell-aspect are 
acceptable (i.e. non-demolition).  Else they can be compensated.
\par 
More generally, for any input state, $w$, of the qubit system, one is 
now in possession of pure Bell-states, which emerge with statistical 
frequencies equal to the theoretical probabilities predicted with that 
$w$, i.e.
$$p_{ab}=\tr _{12}(P_{ab}w)$$ 
To verify this, consider that any mixed state can be written as a convex 
combination of pure states
$$w=\sum\limits_j {\xi _j \ket{\psi _j} \bra{\psi _j}}$$ 
The $\xi _j$ are weights (i.e. $\sum \xi _j=1, \xi _j\ge 0$), and the pure 
states $\ket{\psi _j}$ need not be orthogonal. Any pure state of the qubit 
pair can be expanded in the Bell-basis
$$\ket{\psi } =c_{++}\ket{\Phi ^+} +c_{+-}\ket{\Psi ^+} +c_{-+}\ket{\Phi ^-} +c_{--}\ket{\Psi ^-} $$ 
where
$$\sum\limits_{a,b=\pm} {|c_{ab}|}^2=1$$ 
During the first measurement, there is a projection of $\ket{\psi } $ into 
the subspaces of parallel and anti-parallel spins, and a subsequent bilateral 
rotation by $U_{yy}$. In that fraction of cases, equal to 
$p_-= |c_{+-}|^2+|c_{--}|^2$, where the probe is not flipped, this produces
$$\ket{\psi _{\text{E}} } =\left( {c_{+-}\ket{\Phi ^-} +c_{--}\ket{\Psi ^-} } \right)/ \sqrt {p_-}$$ 
Otherwise, the probe is flipped, and one gets
$$\ket{\psi _{\text{W}} } =\left( {-c_{++}\ket{\Psi ^+} +c_{-+}\ket{\Phi ^+} } \right)/ \sqrt {p_+}$$ 
\begin{figure}
	\centering
	\includegraphics[scale=.8]{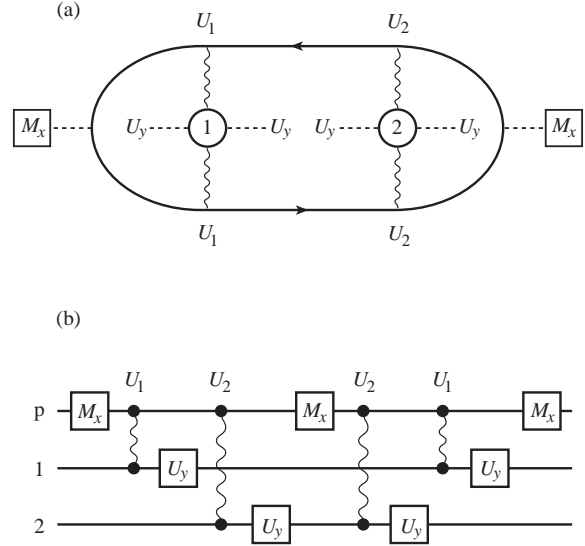}
	\caption{Shuttle-operation in the complete, non-demolition 
	measurement of the Bell-aspect. Equivalent diagrams, for a traveling 
	probe (a), and a stationary probe (b). The wiggly lines represent the 
	interaction (\ref{B}) between the probe p and the qubits. $U_{y}$ are 
	local ${-90}^{\circ}$ qubit rotations, and $M_{x}$ are standard 
	spin~$1/2$ measurements, on the probe exclusively.}
	\label{smallfig}
\end{figure}
The second measurement now gives $\ket{\psi _{\text{EE}}} = \ket{\Psi ^-}$ 
in a fraction of the first cases equal to ${|c_{--}|}^2/ p_-$.
Likewise for the alternative cases. This confirms that one obtains 
the pure output states listed in Table~\ref{table1} with statistical 
frequencies equal to
$$\sum\limits_j {\xi _j {|c_{ab}^{(j)}|}^2}=p_{ab}$$ 
\subsection{Shuttle mode}
All interactions are local. It is therefore possible to operate the 
present procedure in shuttle mode, where the probe can travel between 
qubits located in different regions of space. Of course, an 
instantaneous measurement at space-like separation is not feasible by 
this method. If the rotations $U_y$ can be done at each qubit site, and if 
there is an $M_x$ station, then (\ref{t}) can be arranged as follows
\begin{eqnarray*}
\cdots \underbrace {\left( {U_yU_2} \right)\left( {M_xU_yU_2} \right)}_{\qubit \ 2} &\cdots& \underbrace {\left( {U_yU_1} \right)\left( {\text{init} M_x} \right)}_{\qubit \ 1} \\
&\cdots& \underbrace {\left( {U_yU_1} \right)\left( {M_xU_yU_1} \right)}_{\qubit \ 1} \cdots
\end{eqnarray*}
This is represented graphically in Fig.~\ref{smallfig}.  
\par 
In this shuttle procedure, both of the probe states $\PE$ and $\PW$ 
occur as input, and what is essential in the data is whether the probe 
was flipped or not.  Let {\lq 1\rq} stand for a flip, and {\lq 0\rq} 
for no flip.  There are two types of evolution, provided the shuttle 
action continues.  Either, the state falls into $\ket{\Psi ^-}$ and 
there are no flips of the probe.  Or, there is a cyclic shift among 
the remaining Bell-states
$$\cdots \stackrel{1}{\mapsto} \Phi ^+ \stackrel{1}{\mapsto} \Psi ^+ \stackrel{0}{\mapsto} \Phi ^- \stackrel{1}{\mapsto} \Phi ^+ \stackrel{1}{\mapsto} \Psi ^+ \stackrel{0}{\mapsto} \Phi ^-\cdots $$ 
After the first roundtrip of the probe, provided its initial state is 
known, the measurement is complete. The data uniquely specifies the 
measurement result with respect to the input state. This is what is 
primarily of interest in quantum communication and teleportation. 
\par 
One also knows the specific Bell-state at the output.  Each output 
Bell-state is identified by the three most recent data items, i.e. in 
terms of probe flips
\begin{eqnarray*}
11&:& \Psi ^+ \\
10&:& \Phi ^- \\
01&:& \Phi ^+ \\
00&:& \Psi ^- 
\end{eqnarray*}
In this cycle, after the second roundtrip one has had the opportunity 
to extract any of the three alternatives.  That is, after the two 
first probe measurements one has a pure Bell-state, and subsequently 
every measurement presents the next one.  Therefore, the shuttle 
procedure may be an economical way of preparing specific Bell-states.  
The only statistical element remaining in this preparation procedure, 
even with completely arbitrary input states, for both probe and 
qubits, is whether the system falls into one or the other of the two 
cycles.
\par 
Another interesting possibility is to employ the stability of the 
present shuttle cycle for storage.  Any perturbation, or imperfection, 
will be corrected after one (error-free) roundtrip (2 bits of data).  
With good quality in the operations $U_{1,2}$, $U_y$, and $M_x$, it 
would be expected that imperfections only result in recognizable 
glitches in otherwise perfect cycles.  This would allow persistent 
Zeno monitoring of the Bell-state of the qubit pair.  If the 
probe-qubit interaction is on continuously, what is required is a 
regularly timed sequence of {\lq instantaneous\rq } $M_xU_y$ 
operations.  Such a time-scale gap, apparently, seems to agree with 
the given conditions in some of the relevant systems.
\subsection{Cavity-field probe} \label{subsect:cav}
Suppose a cavity mode interacts dispersively for a specific length of time 
with a passing Rydberg atom 
\cite{Brune92}.  The resulting transformation is of the form
$$U_1=\exp (-i\theta \hat n \otimes  \sigma _z^{(1)})$$
This interaction is suitable for the present purpose, and has the 
advantage that it conserves the photon number.  The number operator 
$\hat n$ generates a phase-shift in the field that occupies the 
cavity.  Thus for two qubit-atoms
\begin{eqnarray*}
U_2U_1&=&\exp (-i2\theta \hat n \otimes  S_z) \\
      &=&1 \otimes  1 \\
	  &&+\bigl( {\cos (2\theta \hat n)-1} \bigr) \otimes P_+-i\sin (2\theta \hat n) \otimes  \szo P_+ \\
      &=&1 \otimes  P_- \\
	  &&+\bigl( {\cos (2\theta \hat n) \otimes  1-i\sin (2\theta \hat n) \otimes  \szo } \bigr)\left( {1 \otimes  P_+} \right)
\end{eqnarray*}
First, consider using cavity number states $\ket{0}$ and $\ket{1}$ 
to create phase-states \cite{Pegg89}
\begin{equation}
\ket{\text{E}} =\roottwo \bigl( { \ket{0} +\ket{1}} \bigr), \quad \ket{\text{W}} =\roottwo \bigl( { \ket{0} -\ket{1}} \bigr)
\label{M}
\end{equation}
To achieve phase-flips between these states the interaction time must 
be adjusted so that $\theta =\pi / 2$. This gives
\begin{equation}
U_2U_1=1 \otimes  P_-+(-1)^{\hat n} \otimes  P_+
\label{L}
\end{equation}
One may then consider passing the qubit-atoms through two such probe 
cavities in succession, with a $U_y$ rotation in between. 
\par 
According to the Jaynes-Cummings model at exact resonance 
\cite{ScullyZubairy97}, the cavity field phase-states (\ref{M}) can be 
transferred to an auxiliary probe-atom, whose state is then measured 
\cite{Haroche95,Brune96a}.  The cavity can be initialized in an 
analogous way \cite{Vogel93}, similar to the one used in 
\cite{Hagley97} to prepare entangled atom states.  Another way to 
measure the phase in two opposite directions has been proposed in 
\cite{Barnett96}.
\par 
The data obtained by measuring the cavity-field phases after both 
atoms have passed would be analogous to those obtained with the 
spin~$1/2$ probe.  With two distinct probe systems, it is possible to 
postpone acquiring the data, since there is no physical interaction 
with the qubits involved in this.  The qubit state is the mixed
$$w''_{\meas}=\sum\limits_{a,b=\pm 1} {P_{ab}wP_{ab}}$$ 
One needs the probe data to separate the four parts in this mixture.  
There is no permutation of the $\Phi $ states, so by including the 
final $U_{yy}^\dagger $ one gets a clean projection on the 
Bell-aspect.  If it is omitted, then there is a permutation 
corresponding to $U_{yy}$.
\par 
Next, consider starting with a coherent state of the probe field
$$\ket{\alpha } =e^{-|\alpha |^2/ 2}\sum\limits_{n=0}^\infty {{{\alpha ^n} \over {\sqrt {n!}}}\ket{n} }$$ 
where $\ket{n}$ are the number eigenstates, and $\alpha $ a complex 
number. For a qubit state  $\ket{\psi }$ one gets
$$U_2U_1\ket{\alpha } \otimes  \ket{\psi } = \ket{\alpha } \otimes  P_-\ket{\psi } +\ket{-\alpha } \otimes  P_+\ket{\psi }$$ 
Although the opposite-phase coherent states are not orthogonal, their 
overlap decreases with field amplitude
$$\langle \alpha |-\alpha \rangle =e^{-2|\alpha |^2}$$ 
These field states resemble classical, i.e. distinct, pointer 
positions.  Furthermore, it has been demonstrated explicitly that the 
coherences in the field density matrix decay in time, more rapidly for 
large amplitudes $|\alpha |$ 
\cite{Brune96a,Davidovich96}.  The time dependence appears to agree with the model 
\cite{Walls85} where the cavity field 
is coupled to a reservoir, which causes amplitude damping: $\alpha \exp (-\lambda t/ 2)$, 
where $\lambda $ is a coupling constant, and $t$ the time.  According to this model, the ratio of off-diagonal 
elements to diagonal ones decreases with time as
$$\biggl( {{{\langle \alpha |-\alpha \rangle } \over {\langle \alpha |\alpha \rangle }}} \biggr)^{1-\exp (-\lambda t)}\cong  e^{-2|\alpha |^2\lambda t}, \quad \text{for} \quad \lambda t\ll 1$$ 
A dispersive coupling allows the qubit-atoms to interact 
simultaneously with the field probe. For moderately large amplitudes 
this could, in principle at least, take place before any significant 
decoherence has occurred. Subsequently, the probe measurement would 
be approximated by the decay of the coherences, still leaving time to 
read the data, before the amplitude reaches zero. Recently a model 
has been proposed \cite{Agarwal98} in which the field amplitude decay 
is prevented, although there is diffusive broadening.
\section{Measuring general entangled aspects}\label{section3}
\subsection{The GHZ-aspect}
Of course, eigenstates of $\sigma _{zz}$, such as $ \ket{\!\uparrow 
\uparrow } $, or $\sigma _{xx}$, such as $ \ket{\! \rightarrow \rightarrow } $, 
are not necessarily entangled.  But the simultaneous eigenstates, the 
Bell-states, are.  For instance
\begin{equation}
\ket{\! \rightarrow \rightarrow } - \ket{\! \leftarrow \leftarrow } 
= \ket{\! \uparrow \downarrow } + \ket{\! \downarrow \uparrow }
\label{o}
\end{equation}
where $ \ket{\! \rightarrow } $ and $ \ket{\! \leftarrow } $ are eigenstates 
of $\sigma _x$.  A three-qubit generalization of (\ref{m}) therefore is
\begin{equation}
P_{abc}=\overeight \bigl( {1+a\sigma _{xxx}} \bigr) \bigl( 
{1+b\sigma _{zz0}} \bigr) \bigl( {1+c\sigma _{0zz}} \bigr)
\label{n}
\end{equation}
with $a,b,c=\pm 1$. The partial projectors commute, since for example
$$\sigma _{xxx}\sigma _{zz0}=\sigma _{xx}\sigma _{zz} \otimes 
 \sigma _x1=\sigma _{zz}\sigma _{xx} \otimes  1\sigma _x=\sigma 
_{zz0}\sigma _{xxx}$$
The 8 entangled states corresponding to $P_{ab}$ (cf. 
Table~\ref{table2}) are mutually orthogonal, since at least one of the 
quantum numbers $a,b,c$ is different between any pair. The set will be 
referred to as the {\lq GHZ-aspect\rq}, since it is built to contain the 
original GHZ-states 
\cite{Greenberger89,Greenberger90,Mermin90} 
$$\roottwo \bigl( { \ket{ \! \uparrow \uparrow \uparrow } \pm  \ket{ \! \downarrow \downarrow \downarrow } } \bigr)$$
Obviously, the GHZ-basis states can be turned into each other by means 
of one or more unilateral spin flips, generated by $\sigma _x$.  
Measurement of this aspect was suggested for the 9-qubit error 
correction code 
\cite{Shor95}.  These states also remain of great 
interest with respect to the purpose for which they were originally proposed.  
\par 
Again, since one starts with just pairwise p/a tests, entanglement 
requires the global e/o test of the projector
\begin{equation}
\bar P_a=\overtwo \bigl( {1+a\sigma _{xxx}} \bigr)
\label{k}
\end{equation}
The complete, non-demolition measurement procedure for the GHZ-aspect 
will be presented in Sect.~\ref{subsect:GHZmeas}. The following is a general 
result, providing modules that perform partial measurements for this and 
other aspects.
\begin{table}
\caption{Basis states of the GHZ-aspect for three qubits.  The two 
representations are equivalent, the second alternative using 
Bell-states of qubits 1 and 2, together with $\sigma _{x}$-eigenstates 
of qubit 3.  \label{table2}}
\begin{tabular}{cccc}
&$abc$&standard basis&${\mathcal{H}}^{(12)} \otimes {\mathcal{H}}^{(3)}$ \\
\tableline					
1&$+++$&$ \ket{ \! \uparrow \uparrow \uparrow } + \ket{ \! \downarrow \downarrow \downarrow } $&$ \ket{\Phi ^+\rightarrow } +\ket{\Phi ^-\leftarrow } $ \\
2&$++-$&$ \ket{ \! \uparrow \uparrow \downarrow } + \ket{ \! \downarrow \downarrow \uparrow } $&$ \ket{\Phi ^+\rightarrow } -\ket{\Phi ^-\leftarrow } $ \\
3&$+-+$&$ \ket{ \! \uparrow \downarrow \downarrow } + \ket{ \! \downarrow \uparrow \uparrow } $&$ \ket{\Psi ^+\rightarrow } -\ket{\Psi ^-\leftarrow } $ \\
4&$+--$&$ \ket{ \! \uparrow \downarrow \uparrow } + \ket{ \! \downarrow \uparrow \downarrow } $&$ \ket{\Psi ^+\rightarrow } +\ket{\Psi ^-\leftarrow } $ \\
5&$-++$&$ \ket{ \! \uparrow \uparrow \uparrow } - \ket{ \! \downarrow \downarrow \downarrow } $&$ \ket{\Phi ^-\rightarrow } +\ket{\Phi ^+\leftarrow } $ \\
6&$-+-$&$ \ket{ \! \uparrow \uparrow \downarrow } - \ket{ \! \downarrow \downarrow \uparrow } $&$ \ket{\Phi ^-\rightarrow } -\ket{\Phi ^+\leftarrow } $ \\
7&$--+$&$ \ket{ \! \uparrow \downarrow \downarrow } - \ket{ \! \downarrow \uparrow \uparrow } $&$ \ket{\Psi ^-\rightarrow } -\ket{\Psi ^+\leftarrow } $ \\
8&$---$&$ \ket{ \! \uparrow \downarrow \uparrow } - \ket{ \! \downarrow \uparrow \downarrow } $&$ \ket{\Psi ^-\rightarrow } +\ket{\Psi ^+\leftarrow } $ 
\end{tabular}
\end{table}
\subsection{The partial measurement module}
The partial measurement corresponding to the generic projector
$$P_a=\overtwo \bigl( {1+a\sigma _{z\cdots z}} \bigr)$$ 
is an operational component in a complete measurement of an $n$-qubit 
entangled aspect.  It is convenient to label the $k$ qubits involved 
($k\le n$) sequentially as $j=1,2,\ldots ,k$.  The remaining $n-k$ 
qubits are ignored during the present procedure.  For each individual 
qubit the quantization axis can be determined by local rotations, 
combined into a multilateral operation, such as $U_{y\cdots y}$.  This 
makes the measurement suitable for testing any $k$-qubit spin 
correlation.  In particular,
$$\bar P_a=U_{y\cdots y}^\dagger P_aU_{y\cdots y}=\overtwo \bigl( {1+a\sigma _{x\cdots x}} \bigr)$$ 
The complete measurement of a GHZ-class entangled aspect can therefore 
be carried out using (essentially) only the present type of partial 
measurement operation. 
\par 
Consider a suitable probe property, $A$, such as for instance $\sigma _z/ 2$ or $\hat n$.  
Suppose the interaction of the probe with the $j$\rq th qubit is
\begin{equation}
U_j=e^{-i\theta A \otimes \sigma _z^{(j)}}=U(\theta ) \otimes P_+^{(j)}+U(-\theta ) \otimes P_-^{(j)} 
\label{2}
\end{equation}
where
\begin{equation}
U(\theta )=e^{-i\theta A},\quad P_\pm ^{(j)}=\overtwo \bigl( {1\pm \sigma _z^{(j)}} \bigr) 
\label{S}
\end{equation}
As usual, $\theta $ is an adjustable angle, determined by the duration 
of the interaction.  It is chosen (since that is adequate) to be the 
same for each $j$.  In order to achieve a measurement it is also 
necessary to choose $A$, and the probe state, carefully.  The probe 
can interact either simultaneously, or sequentially, with the $k$ 
qubits.
\paragraph{Theorem}
After these interactions the joint probe-qubit system has been 
transformed by the following unitary operator
\begin{eqnarray}
U_k&\cdots &U_1 \label{D} \\
&=&U(\alpha _k) \otimes  U_+^{(1\cdots k)}P_+^{(1\cdots k)}+U(\beta _k) \otimes  U_-^{(1\cdots k)}P_-^{(1\cdots k)} \nonumber 
\end{eqnarray}
where
\begin{equation}
\alpha _k=k\pitwo,\quad \beta _k=\alpha _k-\pi 
\label{G}
\end{equation}
The operator to the left of the explicit $\otimes $ is for the probe, 
given by (\ref{S}).  Those to the right are for the $k$ qubits. The projectors 
are ($j=1,\cdots ,k$)
$$P_\pm ^{(1\cdots j)}=\overtwo \bigl( {1^{(1\cdots j)}\pm \sigma _{z\cdots z}^{(1\cdots j)}} \bigr)$$ 
with notation
\begin{eqnarray*}
1^{(1\cdots j)}&\equiv &1 \otimes  \cdots  \otimes  1 \\
\sigma _{z\cdots z}^{(1\cdots j)}&\equiv &\sigma _z^{(1)} \otimes  \cdots \otimes  \sigma _z^{(j)} 
\end{eqnarray*}
The operators $U_\pm $ are given by 
\begin{mathletters}
\begin{eqnarray}
U_+^{(1\cdots j)}&=&U_+^{(1\cdots j-1)}P_+^{(j)}+\eta U_-^{(1\cdots j-1)}P_-^{(j)} \label{Ha} \\
U_-^{(1\cdots j)}&=&U_+^{(1\cdots j-1)}P_-^{(j)}+U_-^{(1\cdots j-1)}P_+^{(j)} \label{Hb} 
\end{eqnarray}
\end{mathletters}
The recursion starts with
$$U_+^{(1)}=U_-^{(1)}=1^{(1)}$$ 
The projectors in these expressions are for a single qubit (the $j$\rq th), 
and $\eta $ is a phase factor to be defined in the 
following.  Note that, if $\eta =1$ this immediately implies that (all $j=1,\cdots ,k$)
$$U_+^{(1\cdots j)}=U_-^{(1\cdots j)}=1^{(1\cdots j)}$$ 
\paragraph{Proof}
The proof is by induction.  As stated in (\ref{2}), for the first 
qubit-probe interaction, $U_1$, the expression agrees in form with (\ref{D}).  
Given that (\ref{D}) is true for the first $j$ qubits, then
\FL\begin{eqnarray}
&&U_{j+1}U_j\cdots U_1 \nonumber \\
&&\enspace =\left\{ {U(\theta ) \otimes P_+^{(j+1)}+U(-\theta ) \otimes P_-^{(j+1)}} \right\} \label{F} \\
&&\quad \times \left\{ {U(\alpha _j) \otimes U_+^{(1\cdots j)}P_+^{(1\cdots j)}+U(\beta _j) \otimes U_-^{(1\cdots j)}P_-^{(1\cdots j)}} \right\} \nonumber 
\end{eqnarray}
(qubit unit operators are not displayed). One needs the identity
\begin{equation}
P_a^{(j+1)}P_b^{(1\cdots j)}=P_a^{(j+1)}P_{ab}^{(1\cdots j+1)} 
\label{E}
\end{equation}
Here, on the left-hand side of (\ref{E})
\FL\begin{eqnarray*}
&&\bigl( {1+a\sigma _z^{(j+1)}} \bigr) \bigl( {1+b\sigma _{z\cdots z}^{(1\cdots j)}} \bigr)  \\
&&\enspace =1+a\sigma _z^{(j+1)}+b\sigma _{z\cdots z}^{(1\cdots j)}\bigl( {a\sigma _z^{(j+1)}} \bigr)^2+ab\sigma _{z\cdots z}^{(1\cdots j+1)} 
\end{eqnarray*}
where a unit operator has been inserted, and this agrees with the 
right-hand side. The four terms in (\ref{F}) are then given by
\begin{eqnarray*}
U(\alpha _j+\theta ) &\otimes&  U_+^{(1\cdots j)}P_+^{(j+1)}P_+^{(1\cdots j+1)} \\
U(\beta _j+\theta ) &\otimes&  U_-^{(1\cdots j)}P_+^{(j+1)}P_-^{(1\cdots j+1)} \\
U(\alpha _j-\theta ) &\otimes&  U_+^{(1\cdots j)}P_-^{(j+1)}P_-^{(1\cdots j+1)} \\
U(\beta _j-\theta ) &\otimes&  U_-^{(1\cdots j)}P_-^{(j+1)}P_+^{(1\cdots j+1)} 
\end{eqnarray*}
If one takes $\theta =\pi / 2$, and if (\ref{G}) holds up to $j$, then 
the parameters in the probe $U$ are related as
\begin{eqnarray*}
\alpha _j+\pitwo&=&\left( {j+1} \right)\pitwo =\alpha _{j+1} \\
\beta _j+\pitwo&=&\left( {j+1} \right)\pitwo-\pi =\beta _{j+1} \\
\alpha _j-\pitwo&=&\left( {j+1} \right)\pitwo-\pi =\beta _{j+1} \\
\beta _j-\pitwo&=&\left( {j+1} \right)\pitwo-2\pi =\alpha _{j+1}-2\pi 
\end{eqnarray*}
The last can give rise to a phase factor
$$\eta =U(-2\pi )=e^{i2\pi A}$$ 
The measurement design requires this to be a number for all relevant 
probe states.  Evidently, in order to obtain a measurement 
transformation, there has to be restrictions of some kind, else one 
would have just any transformation.  For $A=\sigma _z/ 2$ one has 
$\eta =-1$, and for $A=\hat n$ one has $\eta =1$.  If this condition 
(that $\eta $ be a number) is satisfied, then the proof is complete.
\paragraph{Check}
With a spin~$1/2$ probe and $k=2$ one gets
$$\alpha _2=\pi ,\quad \beta _2=0$$ 
and
\begin{eqnarray*}
U_+^{(12)}&=&U_+^{(1)}P_+^{(2)}-U_-^{(1)}P_-^{(2)}=P_+^{(2)}-P_-^{(2)}=\sigma _z^{(2)} \\ 
U_-^{(12)}&=&P_-^{(2)}+P_+^{(2)}=1 
\end{eqnarray*}
So
$$U_2U_1=U(\pi ) \otimes  \sigma _z^{(2)}P_+^{(12)}+U(0) \otimes P_-^{(12)}$$ 
in agreement with (\ref{B}).
\par 
Since the probe advances, or retreats, by $\pi / 2$ in each encounter 
with a qubit, all the four compass states (\ref{l}) can occur.  One gets
\begin{eqnarray*}
\rho '&=&U_k\cdots U_1 \, v \otimes  w \, U_1^\dagger \cdots U_k^\dagger \\
&=&U(\alpha _k) \, v \, U(\alpha _k)^\dagger \otimes U_+^{(1\cdots k)}P_+^{(1\cdots k)}w \, P_+^{(1\cdots k)}{U_+^{(1\cdots k)}}^\dagger \\
&&+U(\alpha _k) \, v \, U(\beta _k)^\dagger \otimes U_+^{(1\cdots k)}P_+^{(1\cdots k)}w \, P_-^{(1\cdots k)}{U_-^{(1\cdots k)}}^\dagger \\
&&+\cdots 
\end{eqnarray*}
With even $k$ one must measure the probe in the x-direction, using 
$M_x$.  For odd $k$ in the y-direction.  Of course, one can then 
consider rotating the probe by $\pi / 2$, in order to be able to use 
the $M_x$ equipment repeatedly.
\par 
For example, with even $k$, starting with $v=\PE$, after the probe has 
been measured, off-diagonal terms in the probe part of $\rho '_{\meas}$ are of the form
$$P_{E,W}U(\alpha _k)\PE U(\alpha _k-\pi )^\dagger P_{E,W}$$ 
They will vanish due to the extra rotation by $\pi $, provided the 
relevant probe states are orthogonal (by design).  Likewise, the 
diagonal terms are
$$P_{E,W}U(\alpha _k) \PE U(\alpha _k)^\dagger P_{E,W}$$ 
and
$$P_{W,E}U(\alpha _k-\pi ) \PE U(\alpha _k-\pi )^\dagger P_{W,E}$$ 
If $k=2s$, and $s=1,3,\ldots $, then the probe state in the first term 
is $\PW$, while in the second term it is $\PE$, and conversely if $s$ 
is even. There remains (odd $s$)
\begin{eqnarray*}
\rho '_{\meas}&=&\PW \otimes U_+^{(1\cdots k)}P_+^{(1\cdots k)}w \, P_+^{(1\cdots k)}{U_+^{(1\cdots k)}}^\dagger  \\
&&+\PE \otimes U_-^{(1\cdots k)}P_-^{(1\cdots k)}w \, P_-^{(1\cdots k)}{U_-^{(1\cdots k)}}^\dagger 
\end{eqnarray*}
and so forth.
\begin{table}
\caption{GHZ-state propagation during the uncompensated measurement 
according to (\ref{n})(overall phase omitted).  The probe starts in 
$\ket{E}$.  Numbers refer to the listing in Table~\ref{table2}, 
letters to the four compass states (\ref{l}) of a qubit probe.  
\label{table3}}
\begin{tabular}{ccccccc}
input&$U_2U_1$&$M_x$&$U_2U_3$&$M_x$&$\bar U_3\bar U_2\bar U_1$&$M_y$ \\
\tableline 
1&5&W&1&E&$\ket{\Psi ^+\rightarrow } -\ket{\Phi ^-\leftarrow } $&S \\
2&6&W&6&W&$\ket{\Phi ^-\rightarrow } -\ket{\Psi ^+\leftarrow } $&S \\
3&3&E&7&W&$\ket{\Psi ^-\rightarrow } -\ket{\Phi ^+\leftarrow } $&S \\
4&4&E&4&E&$\ket{\Phi ^+\rightarrow } -\ket{\Psi ^-\leftarrow } $&S \\
5&1&W&5&E&$\ket{\Phi ^-\rightarrow } +\ket{\Psi ^+\leftarrow } $&N \\
6&2&W&2&W&$\ket{\Psi ^+\rightarrow } +\ket{\Phi ^-\leftarrow } $&N \\
7&7&E&3&W&$\ket{\Phi ^+\rightarrow } +\ket{\Psi ^-\leftarrow } $&N \\
8&8&E&8&E&$\ket{\Psi ^-\rightarrow } +\ket{\Phi ^+\leftarrow } $&N 
\end{tabular}
\end{table}
\subsection{Compensation}\label{subsect:comp}
Besides projections by $P_{\pm}$, the $k$-qubit system may experience some unitary 
transformations, as when $\eta =-1$ for the qubit probe.  One has
\begin{eqnarray*}
U_+^{(1\cdots k)}{U_+^{(1\cdots k)}}^\dagger &=&U_+^{(1\cdots k-1)}{U_+^{(1\cdots k-1)}}^\dagger P_+^{(k)} \\
&&+{|\eta |}^2 \, U_-^{(1\cdots k-1)}{U_-^{(1\cdots k-1)}}^\dagger P_-^{(k)} \\
&=&P_+^{(k)}+P_-^{(k)}=1 
\end{eqnarray*}
using the recursion (\ref{Ha}), and likewise for $U_{-}$.  It also follows that $U_\pm $ and $P_\pm 
$ commute.  The unitary operators $U_\pm $ are Hermitean, by recursive 
construction.  Consequently
$${\bigl( {U_\pm ^{(1\cdots k)}} \bigr)}^2=1^{(1\cdots k)}$$ 
Therefore, if the unitary transformations $U_\pm $ on the selected $k$ 
qubits are not desirable, a clean projection can be obtained by 
repeating the measurement, as described already for the Bell-aspect in 
(\ref{I}).
\par 
For the qubit probe case, there is an alternative way to compensate 
the unitary operations.  For the $k$ qubits, let
$$U_{z\cdots z}^{(1\cdots k)}=U_z^{(1)} \otimes  \cdots  \otimes  U_z^{(k)}$$ 
where
$$U_z=e^{-i{\pitwo }\sigma _z/ 2}=\roottwo \bigl( {1-i\sigma _z} \bigr) $$ 
This represents a multilateral $+\pi / 2$ rotation about the z-axis.  
Then it is straightforward to show that
\FL\begin{eqnarray}
&&\enspace U_{z\cdots z}^{(1\cdots k)}U_k\cdots U_1 \nonumber \\
&&\quad =U(\alpha _k) \otimes g^kP_+^{(1\cdots k)}+U(\beta _k) \otimes  ig^kP_-^{(1\cdots k)} 
\label{K}
\end{eqnarray}
with 
$$g=e^{-i\pi / 4}=\roottwo \left( {1-i} \right) $$ 
For instance, one has
$$U_zP_+=gP_+ , \quad U_zP_-=igP_- $$ 
\par 
Another way is to write
$$\tilde U_j=U_z^{(j)}U_j=e^{-i\theta {\overtwo }(1+\sigma _z) \otimes  \sigma _z^{(j)}}$$ 
and use it for $\theta =\pi / 2$ as before, but with $A=P_+$ for the probe, i.e. 
$$\tilde U(\theta )=e^{-i\theta P_+}=1+\left( {e^{-i\theta }-1} \right) P_+$$ 
This gives $\eta =\tilde U(-2\pi )=1$.  In the expression (\ref{D}) one 
must then use $U_\pm =1$, together with
\begin{eqnarray*}
\tilde U(\alpha _k)&=&e^{-ik{\pitwo }P_+}=g^kU(\alpha _k) \\ 
\tilde U(\beta _k)&=&e^{-i(k-2){\pitwo }P_+}=ig^kU(\beta _k) 
\end{eqnarray*}
Apart from the phase factors, the present global rotation compensates 
as effectively as repeating the measurement, although it requires a 
different set of operations from what is otherwise used, i.e. the 
$U_{z}$.  In some systems, this rotation could perhaps be drawn from 
an inherent {\lq precession\rq} of the qubits (not included here).
\subsection{Shuttle-measurement of the GHZ-states}\label{subsect:GHZmeas}
Entangled states with a structure such as the one shown in (\ref{n}) 
for the eight GHZ-basis states can be measured in a complete, 
non-demolition fashion by means of the partial measurement module just 
established.  Of course, these modules can also be used separately in 
ways, which need not imply entanglement.  In each stage, there are 
options to repeat or to compensate, as described above in 
Sect.~\ref{subsect:comp}.  For the following design, it will be 
regarded as acceptable that there are simple cycles between the 
members of the GHZ-basis.
\begin{table}
\caption{GHZ-state propagation during two rounds of shuttle operation 
(overall phase omitted).  The probe starts (and ends) in $\ket{E}$.  
Numbers refer to the listing in Table~\ref{table2}.  \label{table4}}
\begin{tabular}{cccc}
input&first-round data&output&second-round data \\
\tableline 
1&WESW&5&EWSE \\
2&WWSE&2&WWSE \\
3&EWSE&3&EWSE \\
4&EESW&8&WWSE \\
5&WENW&1&EWNE \\
6&WWNE&6&WWNE \\
7&EWNE&7&EWNE \\
8&EENW&4&WWNE
\end{tabular}
\end{table}
\par 
The probe is initialized in the state $\PE$.  It first interacts with 
qubits 1 and 2, by $U_1$ and $U_2$, and is then measured in the 
x-direction, $M_x$.  Secondly, it interacts with qubit 3 and again 
with qubit 2, via $U_3$ and $U_2$.  It is then measured again with 
$M_x$.  When the GHZ-basis states are given as input, this 
non-demolition processing leads to some permutations, as listed in 
Table~\ref{table3}.  They come from
\begin{eqnarray*}
U_+^{(12)}&=&P_+^{(2)}-P_-^{(2)}=\sigma _z^{(2)},\quad U_-^{(12)}=P_-^{(2)}+P_+^{(2)}=1 \\
U_+^{(32)}&=&P_+^{(2)}-P_-^{(2)}=\sigma _z^{(2)},\quad U_-^{(32)}=P_-^{(2)}+P_+^{(2)}=1 
\end{eqnarray*}
The pattern of probe data (2 bits) now allows to distinguish four 
orthogonal subspaces, but there is not yet any certainty of 
entanglement.
\par 
The e/o test starts by tilting each qubit spin by means of $U_y$.  
Then all qubits interact with the probe.  Returning the qubits by 
$U_y^\dagger $, all in all (overbars indicate that the operators are 
built from $\sigma _x$ instead of $\sigma _z$)
\begin{eqnarray*}
\bar U_3\bar U_2\bar U_1&=&U_{yyy}^ \dagger U_3U_2U_1U_{yyy} \\
&=&\left( {U_y^ \dagger U_3U_y} \right)\left( {U_y^ \dagger U_2U_y} \right)\left( {U_y^ \dagger U_1U_y} \right) 
\end{eqnarray*}
Finally the probe is measured in the y-direction, $M_y$.  Or, it is 
rotated by $-\pi / 2$, say, and measured with the $M_x$ equipment 
(this option will not be included here).  According to (\ref{D}) the 
interactions produce
\begin{eqnarray*}
&&\bar U_3\bar U_2\bar U_1 \\
&&\enspace =U(3\pi / 2) \otimes  \bar U_+^{(123)}\bar P_+^{(123)}+U(\pi / 2) \otimes  \bar U_-^{(123)}\bar P_-^{(123)} 
\end{eqnarray*}
Here
$$\bar P_\pm ^{(123)}=\overtwo  \bigl( {1^{(123)}\pm \sigma _{xxx}^{(123)}} \bigr)$$ 
As the probe starts in $\PE$, say, when operating on the GHZ-states it 
will become either $\PN$ or $\PS$, depending on the outcome of the e/o 
test, and hence must be measured in the y-direction.  With $\eta =-1$, 
the unitary operators are
$$\bar U_+^{(123)}=\sigma _x^{(2)}\bar P_+^{(3)}-\bar P_-^{(3)}, \quad  \bar U_-^{(123)}=\sigma _x^{(2)}\bar P_-^{(3)}+\bar P_+^{(3)}$$ 
These determine which 3-qubit states are available after completion 
of the measurement. Use the following representation
$$\ket{\Psi \rightarrow } ^{(123)}\equiv \ket{\Psi } ^{(12)}\otimes \ket{ \rightarrow } ^{(3)}$$ 
The first factor in the right-hand expression stands for any of the 
Bell-states, so
$$\sigma _{0x}\ket{\Psi ^\pm }  = \ket{\Phi ^\pm } ,\quad \sigma _{0x}\ket{\Phi ^\pm }  = \ket{\Psi ^\pm } $$ 
and the last factor is a $\sigma _x$ eigenstate.  The GHZ-basis in 
this representation is shown in Table~\ref{table2}.  As usual, the 
even states 1 to 4 are operated upon by $\bar U_+$, while the probe 
turns $3\pi / 2$, and the odd states 5 to 8 by $\bar U_-$, the probe 
turning $\pi / 2$.
\par 
The resulting qubit output states and probe data are listed in 
Table~\ref{table3}.  It depends, of course, on the circumstances, 
whether these output states are useful.  In any event, there is a 
compensation procedure using only local rotations, $U_z$, which will 
produce output within the original GHZ-aspect.  Another such 
alternative output-aspect is entered if one omits the final 
$U_{yyy}^{\dagger }$.
\begin{figure}
	\centering
	\includegraphics[scale=.8]{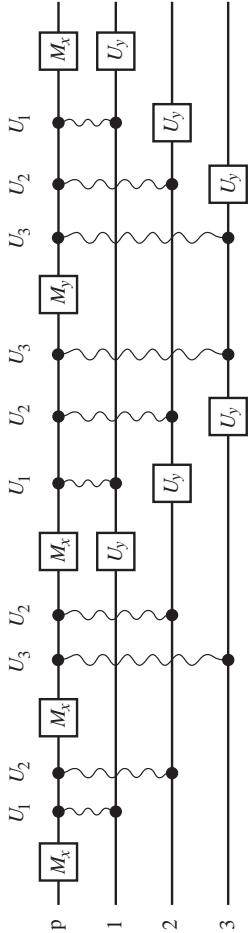}
	\caption{Shuttle-operation of the complete, non-demolition 
	measurement of the GHZ-aspect. The probe p interacts with qubits 1, 
	2, and 3 via (\ref{2}) (wiggly lines) and is measured in the x- and 
	y-directions at the standard measuring stations $M_{x}$ and $M_{y}$. 
	The qubits are rotated by local $U_{y}$. The GHZ-states pass 
	unhindered, except for some permutations.}
	\label{bigfig}
\end{figure}
\par 
Repeating the last partial measurement corresponds to doing
\begin{eqnarray*}
&&\left( {M_x\bar U_1\bar U_2\bar U_3} \right) \left( {M_y\bar U_3\bar U_2\bar U_1} \right) \\
&&\enspace =\left( {M_xU_{yyy}^ \dagger U_1U_2U_3U_{yyy}} \right)\left( {M_yU_{yyy}^ \dagger U_3U_2U_1U_{yyy}} \right) 
\end{eqnarray*}
According to Sect.~\ref{subsect:comp}, this will return the appropriate 
GHZ-states, i.e. the ones that are present after the first two stages 
in Table~\ref{table3}.  
\par 
This allows the following shuttle design.  Cancel the intermediate 
rotations by the unitarity of $U_{yyy}$.  At the end, add $U_{yyy}^2=-i\sigma _{yyy}$, 
which merely performs another permutation on the GHZ-basis (exchanging the even and odd states, i.e. 
(15)(26)(37)(48) apart from phase factors $\pm 1$).  This gives the 
shuttle procedure of Fig.~\ref{bigfig}
$$\left( {M_xU_{yyy}U_1U_2U_3} \right)\left( {M_yU_3U_2U_1U_{yyy}} \right) $$ 
The final output is shown in Table~\ref{table4}.  One must of course 
do the third probe measurement, $M_y$, in order to complete the projection of the qubit 
system into one-dimensional subspaces. The aspect at this stage is not 
GHZ. The data from the fourth probe measurement, $M_x$, is 
redundant, but a useful check on the e/o test. The unitary operations 
preceding it produce the GHZ output-states listed.
\par 
Having returned to the GHZ-aspect, the measurement 
sequence can now be repeated. Apart from 
phase-factors, this merely gives rise to the simple cycles
$$\cdots 1\mapsto 5\mapsto 1\cdots , \quad  \cdots 4\mapsto 8\mapsto 4\cdots $$ 
After two rounds, therefore, all the GHZ-basis states emerge in the 
original places, and the probe is in $\PE$, where it started.  The 
GHZ-states also result when the input is an arbitrary mixed state, 
$w$, with statistical frequencies equal to the predicted probabilities 
for $w$.  One can recognize the outcome of the complete, 
non-demolition measurement by the probe data sequence that has been 
recorded.  For two rounds, as shown in Table~\ref{table4}, only eight 
8-bit words out of a total of 256 possible ones correspond to an 
error-free measurement.  These words are going to repeat as the 
shuttle continues running.
\section{Conclusions}\label{section4}
The measurement procedure established in the present paper is different from the 
existing ones in at least one of the following respects: (a) It is a 
complete, non-demolition measurement process, under which all the 
entangled basis states are invariant.  (b) The entire interaction with 
the entangled system is taken care of by a single (qubit) probe, used 
repeatedly.  (c) There is no interaction between the individual parts 
of the entangled system.  (d) There is no interaction between the 
environment and the entangled system, only with the probe.  (e) All 
interactions are two-qubit interactions between the probe and each of 
the constituents of the entangled system in turn.  (f) This 
interaction is of the simplest conceivable form, and only one kind of 
probe-qubit interaction is required to perform the entire measurement.  
(g) The design is modular, so that it can perform partial 
measurements, and can be extended to handle entangled states of many 
qubits.  (h) The same procedure can be adjusted to measure a wide 
range of different entangled states by local qubit operations, such as 
unilateral spin rotations.
\par 
Apart from this one needs external apparatus, which can perform
measurements on the probe. It is not essential how the apparatus is 
built, as long as it performs a standard measurement, 
say measures the probe spin in the x-direction. This equipment 
can also be localized, i.e. it does not require entangled states of 
the apparatus to be prepared and distributed beforehand.
\par 
With the present procedure the measurement process is split up into 
exclusively local operations and interactions, which can take place 
sequentially as the probe visits each of the qubit components in turn.
All these operations are elementary, and in their totality they can be 
said to define the operational nature of the observable entanglement.
\par 
It is possible to design the probe schedule in a shuttle mode, which 
may be repeated indefinitely.  In this way one can take advantage of 
the quantum Zeno effect to maintain the entangled states for a length 
of time, with an insignificant probability of decoherence.  Such a 
device would be capable, in principle at least, to store entangled 
states until they are released at a predictable time, in a predictable 
state.  The shuttle process is stable, and self-correcting, in the 
sense that any random errors are removed after a few steps by the 
probe measurement actions.

\end{document}